\documentstyle[12pt,epsfig]{article}
\def\beq{\begin{equation}}
\def\eeq{\end{equation}}
\def\bea{\begin{eqnarray}}
\def\eea{\end{eqnarray}}

\begin{document}

\begin{center}
{\Large \bf  Various Scattering Properties of a New PT-symmetric non-Hermitian potential  
  }

\vspace{1.3cm}

{\sf Ananya Ghatak \footnote{e-mail address: gananya04@gmail.com}
Raka Dona Ray Mandal \footnote{e-mail address: rakad.ray@gmail.com}
and Bhabani Prasad Mandal \footnote{e-mail address:
\ \  bhabani.mandal@gmail.com,\ \ bhabani@bhu.ac.in  }}

\bigskip

{\em  Department of Physics,
Banaras Hindu University, Varanasi-221005, INDIA.\\
$\dagger$ Department of Physics,
Rajghat Besant School,
Varanasi-221001, INDIA. \\
}

\bigskip
\bigskip

\noindent {\bf Abstract}
\end{center}
We complexify a 1-d potential
$V(x)=V_0 \cosh ^2\mu \{\tanh[(x-\mu d)/d]+\tanh(\mu)\}^2$
which exhibits bound, reflecting and free states to study various properties of a
non-Hermitian system. This potential turns out a PT-symmetric non-Hermitian potential when
one of the parameters $(\mu, d)$ becomes imaginary. For the case of $\mu\rightarrow i\mu$, we 
have entire 
real bound state spectrum. Explicit scattering states are constructed to show reciprocity at 
certain discrete values of energy even though the potential is not parity symmetric. 
Coexistence of deep energy minima of transmissivity with the multiple spectral singularities 
(MSS) is observed. 
We further show that this potential becomes invisible from left (or right) at certain
discrete energies. The penetrating states in the other case ($d\rightarrow i d$) are always 
reciprocal even though it is PT-invariant and no spectral singularity (SS) is present in this 
case. Presence of MSS and reflectionlessness are also discussed 
for the free states in the later case.

\medskip
\vspace{1in}
\newpage
\section{Introduction}
PT-symmetric non-Hermitian system \cite{ ben4} with real energies are very exciting, because 
one can have fully consistent quantum theory by restoring the Hermiticity and
by upholding the unitary time evolution for such systems in a modified Hilbert space
\cite{mos, ben}. These interesting results help this subject to grow exponentially and to find
many applications in different branches of physics \cite{ qin}-\cite{opt3}. Experimental 
realizations of such systems in optics have further enhanced the motivations for such studies
\cite{ opt1, opt2, opt3}. However not all PT-symmetric non-Hermitian systems lead to fully consistent
quantum theory because of the presence of exceptional points (EP) \cite{ep0, ep1, ep2} and spectral singularities
(SS) \cite{ss1, ss2, ss3} where two or more eigenvalues along with corresponding eigen functions coalesce. 
This phenomenon is known to have physically observable consequences. Scattering 
cross-section and/ or reflection and transmission current diverges. Thus it is extremely
important to study the presence of such singular points in non-Hermitian systems.

For any real or complex potential, reflectivity ($R$) and transmissivity ($T$) do not depend on 
the direction of incidence as long as the potential remains invariant under parity 
transformation. PT-symmetric non-Hermitian potentials are naturally parity non-invariant 
and hence the reflection coefficients for left ($R_l$) incident particle and right ($R_r$) incident particle
are generally not equal to each other, i.e. $R_{l}\not=R_{r}$ , even though $T_{l}=T_{r}=T$. Thus all the PT-symmetric 
non-Hermitian potentials are non-reciprocal. In other word they show handedness  \cite{ss3}. 
However at certain 
discrete energies these PT-symmetric non-Hermitian potentials become reciprocal, i.e.   $R_{l}=R_{r}$
and/or they become unitary ($R$+$T$=1). It is extremely important to find such energy points 
to get further insight into these systems. 

Invisibility of a complex potential is another important aspect of scattering \cite{inv2, inv1, inv3}. 
The potential is called invisible from left if $R_l=0$, 
$R_r\not=0$ and in addition $T$=1; similarly the potential become invisible from right if 
$R_r=0$, $R_l\not=0$ and $T$=1. Recently it has been shown that the 
equations governing invisibility of the potential are invariant under combined parity and
time reversal transformation \cite{inv1}. However PT-symmetry is neither necessary nor 
sufficient
for the invisibility of a scattering potential. The concept of invisibility could lead to
many applications in optics where analogy is made between complex potential and complex 
refractive index of a optical material. The efficiency of optical device is enhanced by  
reducing the level of absorption \cite{opap, opapp}.

The purpose of the present work is to study various aspect of these scattering properties \cite{lev, ka}
by considering a new PT-symmetric non-Hermitian potential.
We consider a 1-d real potential \cite{mors}
\beq
V(x)=V_0 \cosh ^2\mu \{\tanh[(x-\mu d)/d]+\tanh(\mu)\}^2
\label{real}
\eeq
which admits bound, reflecting and free state solutions depending on the energy of the 
particle.
$V_0$ is the depth of the potential, the parameter d controls the width of the potential
 and $\mu$ decides whether
the potential behaves like a well or a barrier. We complexify this potential in 
PT-symmetric manner by allowing the
parameters ($\mu, d$) to become imaginary separately. We show that entire bound state spectrum is
real and the system always remains in unbroken phase of PT-symmetry when $\mu\rightarrow i\mu$.
By considering explicit scattering states we show that reciprocity is 
restored even for a PT-symmetric non-Hermitian system at certain discrete values of energy.
This system has multiple spectral singularities which coexist with deep energy minima points
of the transmission coefficient.
The potential becomes invisible from left (or right) at different energy values.
It is possible to make this potential bidirectional invisible at several energies 
by choosing appropriate values of the parameters in the potential.

In the other case ($d\rightarrow i d$) we have penetrating as well as free states. 
Remarkably penetrating states are reciprocal for all energies even though the system is 
PT-symmetric non-Hermitian. No SS is present in this case. However MSS are shown
to be present for the free states. Aspects of invisibility are also discussed for this case.

Now we present the plan of this paper. In Sec. 2 we discuss the nature of the real potential 
and complexify it in a PT-symmetric manner. In Sec. 3.1 we discuss the bound states of the 
non-Hermitian potential when $\mu\rightarrow i\mu$. Various aspects of scattering states
for this case have been considered in Sec. 3.2. Sec. 4 is for the discussion of 
penetrating and free states for the second case ($d\rightarrow i d$). Sec. 5 is reserved for 
discussions and conclusions.

\section{The Potential and PT-symmetrization}

The potential in Eq. (\ref{real}) has a minima at x=0 and asymptotically goes to a finite
value $V_0e^{\pm 2\mu}$ as $x\rightarrow \pm \infty$. This potential exhibits bound states, 
reflective states and 
free states depending on the energy of the incident particle. The particle have bound states
if its energy is between 0 to $V_0e^{-2\mu}$ (region I in the Fig.1) and it oscillates
back and fourth inside the potential. If the particle energy E is in between 
$V_0e^{-2\mu}$ and $V_0e^{2\mu}$ then we have reflecting states (region II in the Fig.1). 
Particle comes
from left reflected back to $-\infty$ by the potential rise to the right. In the region 
III of the Fig.1 particle can move freely between $-\infty$ to $+\infty$ as its energy
$E>V_0e^{2\mu}$. \\

\ \ \ \ \ \ \ \ \ \ \ \ \ \ \ \ \ \ \ \ \ \includegraphics[scale=.9]{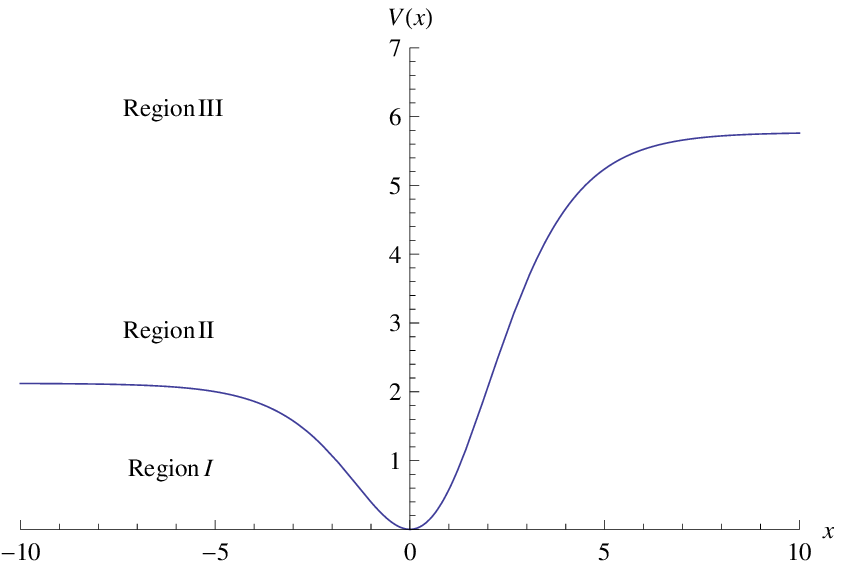} \\

{\bf Fig.1}: {\it 1-d real potential ($V_0 = 3.5, \mu = .25$) which admits bound (Region-I),
reflection (Region-II) and free states (Region-III).} \\

This Hermitian potential in Eq. (\ref{real})
can be complexified in many ways by considering one or more of the parameters $\mu$, $d$ 
and $V_0$ complex.
However not all of them are PT-symmetric which are of our interest. In two simple ways this 
potential can be made PT-symmetric non-Hermitian;

{\bf Case-I.}
When $\mu$ is purely imaginary ($\mu\rightarrow i\mu$), the potential becomes,
\beq
V(x)=V_0 \cosh ^2(i\mu) \{\tanh[(x-i\mu d)/d]+\tanh(i\mu)\}^2
\label{pt1}
\eeq
This potential is invariant under combined PT-transformation.

{\bf Case-II.}
On the other hand the potential in Eq. (\ref{real}) turns out to be a PT-symmetric 
non-Hermitian 
if we substitute $d\rightarrow id$ with a purely imaginary shift $\zeta $ (where $\zeta$ can 
have arbitrary values between $-\infty$ to $+\infty $)
to the x coordinate and written as,
\beq
V(x+i \zeta )=V_0 \cosh ^2\mu \{\tanh[(x+i \zeta  -\mu i d)/id]+\tanh(\mu)\}^2
\label{pt2}
\eeq
In the first case we have entirely real bound states and PT-symmetry is unbroken for all
values of the parameter. Scattering states show many important features. In the second case we have penetrating states and free states.

\section{Non-Hermitian PT-symmetric: case I ($\mu\rightarrow i\mu$)}
We begin the section with the general solution of the Schroedinger equation 
for the Hermitian potential in Eq. (\ref {real}) which is given by,
\beq
\psi (z)=\frac{e^{-az}}{{(e^z+e^{-z})}^b} \left[N F_1+M\left(\frac{e^{z}}{{e^z+e^{-z}}}\right)^
{-(a+b)} F_2\right]
\label{hr}
\eeq
where,
$z=(x-\mu d)/d$, $v=(2md^2/\hbar^2)V_0$, $\epsilon=(2md^2/\hbar^2)E$, M and N are the
normalization constants. $F_1$ and $F_2$ are the two solutions of hypergeometric equation,
\bea
F_1&=&F\left(b+1/2-\sqrt{v\cosh^2 \mu +1/4},
 b+1/2+\sqrt{v\cosh^2 \mu +1/4};\mid a+b+1\mid ;\frac
{e^{-z}}{e^z+e^{-z}}\right) \nonumber \\
F_2&=&F\left(-a+1/2-\sqrt{v\cosh^2 \mu +1/4},
 -a+1/2+\sqrt{v\cosh^2 \mu +1/4};\mid -a-b+1\mid ;\frac
{e^{-z}}{e^z+e^{-z}}\right) \nonumber \\
\label{ff}
\eea
where the parameters $a$ and $b$ are defined as,
\beq
a^2+b^2=-\epsilon +v\cosh 2\mu; 2ab=v\sinh2\mu .
\label{ab}
\eeq
Eq. (\ref{ab}) further can be expressed as,
\bea
a&=&\frac{1}{2}\sqrt{ve^{2\mu}-\epsilon }-\frac{1}{2}\sqrt{ve^{-2\mu}-\epsilon }
\equiv \frac{1}{2}\kappa _+-\frac{1}{2}\kappa_- ; \nonumber \\
b&=&\frac{1}{2}\sqrt{ve^{2\mu}-\epsilon }+\frac{1}{2}\sqrt{ve^{-2\mu}-\epsilon }
\equiv \frac{1}{2}\kappa _++\frac{1}{2}\kappa_- ; \nonumber \\
\label{oab}
\eea
As for bound states $\epsilon $ is always less than $ve^{\pm 2\mu}$, $\kappa_+$ and $\kappa_-$ 
are real for the Hermitian case.
The solution of the Schroedinger equation which is finite at $z=\infty $ is
therefore obtained from the general solutions in Eq. (\ref{hr}) as,
\bea
\psi &=& \frac{e^{-az}}{{(e^z+e^{-z})}^b} N F_1=\frac{Ne^{-az}}{{(e^z+e^{-z})}^b} F\left(b+1/2-\sqrt{v\cosh^2 \mu +1/4},\right. 
\nonumber \\
 && \left. \ \ \ \ \ \ b+1/2+\sqrt{v\cosh^2 \mu +1/4};\mid a+b+1\mid ;\frac
{e^{-z}}{e^z+e^{-z}}
\right) \nonumber \\
\eea

which has the following limiting behavior at $z\rightarrow - \infty $,
\bea
\psi &\rightarrow & \frac{\Gamma(a+b+1)\Gamma(b-a)e^{(a-b)z}}{\Gamma(b+1/2-\sqrt{v
\cosh^2 \mu +1/4})\Gamma(b+1/2+\sqrt{v\cosh^2 \mu +1/4})} \nonumber \\
&& \ \ \ \ \ \ \ +  \frac{\Gamma(a+b+1)\Gamma(a-b)e^{(b-a)z}}{\Gamma(a+1/2-\sqrt{v
\cosh^2 \mu +1/4})\Gamma(a+1/2+\sqrt{v\cosh^2 \mu +1/4})} \nonumber \\
\eea 
The allowed bound states are obtained for the finite asymptotic wavefunction depending on 
the situations $a>b$ or $b>a$.

Now with this general discussion on the Hermitian potential we are ready to take up
the non-Hermitian extensions of this potential.

\subsection{Bound States}
In this case we denote,
$z=(x-i\mu d)/d$ and re-express $a$ and $b$ by using Eq. (\ref{ab}) for 
an imaginary $\mu$ as,
\bea
a^I&=&\frac{1}{2}\sqrt{ve^{2i\mu}-\epsilon }-\frac{1}{2}\sqrt{ve^{-2i\mu}-\epsilon }
\equiv \frac{1}{2}k _+-\frac{1}{2}k_- ; \nonumber \\
b^I&=&\frac{1}{2}\sqrt{ve^{2i\mu}-\epsilon }+\frac{1}{2}\sqrt{ve^{-2i\mu}-\epsilon }
\equiv \frac{1}{2}k_++\frac{1}{2}k_- . \nonumber \\
\eea
From the above equation one can easily see that $a^I$ becomes purely imaginary and on the other hand 
$b^I$ is still real.
The general solution for the Schroedinger equation for this non-Hermitian PT-symmetric 
potential in Eq. (\ref {pt1}) now is written (with $a^I\rightarrow ia'$) ,
\beq
\psi^I(z)=\frac{Ne^{-ia'z}}{{(e^z+e^{-z})}^{b^I}} F_1^I+M\left(\frac{e^{z}}{{e^z+e^{-z}}}\right)^
{-(ia'+b^I)}\frac{e^{ia'z}}{{(e^z+e^{-z})}^{b^I}} F_2^I
\label{gn}
\eeq
where $F_1^I$ and  $F_2^I$ are obtained from Eq. (\ref{ff}) by substituting $a, b$ and $z$ 
for the case of imaginary $\mu$.
The finite wavefunction at $z\rightarrow +\infty$ for the bound states is,
\bea
\psi^I(z) &=& \frac{Ne^{-ia'z}}{{(e^z+e^{-z})}^{b^I}} F\left(b^I+1/2-\sqrt{v\cosh^2 (i\mu) +1/4},\right. 
\nonumber \\
 && \left. \ \ \ \ \ \ b^I+1/2+\sqrt{v\cosh^2 (i\mu) +1/4};\mid ia'+b^I+1\mid ;\frac
{e^{-z}}{e^z+e^{-z}}
\right) \nonumber \\
\label{ptsi}
\eea

The above wave function has the following limiting behavior at $z\rightarrow - \infty $,
\bea
\psi^I(z) &\rightarrow & \frac{\Gamma(ia'+b^I+1)\Gamma(b^I-ia')e^{ia'z}e^{-b^Iz}}{\Gamma(b^I+1/2-\sqrt{v
\cosh^2 (i\mu) +1/4})\Gamma(b^I+1/2+\sqrt{v\cosh^2 (i\mu) +1/4})} \nonumber \\
&& \ \ \ \ \ \ \ +  \frac{\Gamma(ia'+b^I+1)\Gamma(ia'-b^I)e^{-ia'z}e^{b^Iz}}{\Gamma(ia'+1/2-\sqrt{v
\cosh^2 (i\mu) +1/4})\Gamma(ia'+1/2+\sqrt{v\cosh^2 (i\mu) +1/4})} . \nonumber \\
\eea 
For $b^I>0$, to have a finite wavefunction at $z\rightarrow - \infty $ the argument 
of one of the Gamma function in the denominator of the first
term  must be a negative integer. This fixes the  
allowed ranges of energy for the bound states where,
\beq
b^I=b^I_n=\sqrt{v\cos^2 \mu +1/4}-(n+1/2); \ \ \ a^I=a^I_n=\frac{i v \sin 2\mu}
{2\left(\sqrt{v\cos^2 \mu +1/4}-(n+1/2)\right)} .
\label{con}
\eeq
The bound state energy for this case is then $\epsilon^I_n = v\cos (2\mu)-(a^I)^2-(b^I)^2$,
\beq
\epsilon^I_n = v\cos (2\mu)+ \frac{v^2\sin^2 2\mu}{4\left(
\sqrt{v\cos^2 \mu +1/4}-(n+1/2)\right)^2}-[\sqrt{v\cos^2 \mu +1/4}-(n+1/2)]^2
\label{e1}
\eeq
where, $n=0,1,2...<(\sqrt{v\cos^2 \mu +1/4}-1/2)$, and the corresponding wave function is,
\beq
\psi^I_n = \frac{Ne^{-a^I_nz}}{(e^z+e^{-z})^{b^I_n}} F\left(-n,2\sqrt{v\cos^2 \mu +1/4}-n;\mid 
a^I_n+b^I_n+1\mid ;\frac {e^{-z}}{e^z+e^{-z}}\right) .
\label{n1}
\eeq
Now to have a PT-unbroken phase the wavefunction need to satisfy the condition 
$PT\psi =A \psi$. The PT-transformation changes $a_n\rightarrow -a^I_n, z\rightarrow -z, b_n\rightarrow b^I_n, v\rightarrow v$ and hence we obtain,
\beq
PT\psi^I_n=\frac{Ne^{-a^I_nz}}{(e^z+e^{-z})^{b^I_n}} F\left(-n,2\sqrt{v\cos^2 \mu +1/4}-n;\mid 
-a^I_n+b^I_n+1\mid ;\frac {e^{z}}{e^z+e^{-z}}\right) 
\label{n1}
\eeq
Using the standard properties of hypergeometric functions we obtain $PT\psi^I_n$ 
for the case of bound states as,
\bea
PT\psi^I_n &=&\frac{\Gamma(a^I_n+b^I_n+1)\Gamma(a^I_n-b_n^I)}
{\Gamma (a^I_n+b^I_n+n+1)\Gamma (a^I_n-b^I_n-n)}\left(\frac{Ne^{-a_n^Iz}}{{(e^z+e^{-z})}^{b_n^I}} \right) \nonumber \\
&&
F\left(-n,2\sqrt{v\cos^2 \mu +1/4}-n;\mid 
a^I_n+b^I_n+1\mid ;\frac {e^{-z}}{e^z+e^{-z}}\right) =A_n\psi^I_n \nonumber \\
\eea
where, $$A_n =\frac{\Gamma(a^I_n+b^I_n+1)\Gamma(a^I_n-b_n^I)}
{\Gamma (a^I_n+b^I_n+n+1)\Gamma (a^I_n-b^I_n-n)} \ \mbox {\ \ is a pure number for

the n-th state.}$$
In this case the entire bound state spectrum is in PT-unbroken phase 
and the energy eigenvalues in Eq. (\ref{e1}) is always real. Similar result can also be obtained
for $b^I<0$.

\subsection{Scattering States and Spectral Properties}
Now for the scattering states we have to consider all the 
solutions of Schroedinger equation for the PT symmetric potential in Eq. (\ref{pt1}),
\bea
\psi_1 (z) &=& Ne^{\frac{1}{2}i(k_+-k_-)z}(e^z+e^{-z})^{\frac{1}{2}i(k_++k_-)z} \nonumber \\
&.& F\left(-\frac{1}{2}ik_+-\frac{1}{2}ik_-+\frac{1}{2}-\gamma,-\frac{1}{2}ik_+-\frac{1}{2}ik_-
+\frac{1}{2}+\gamma ;\mid 1-ik_+\mid;\frac{e^{-z}}{e^z+e^{-z}}\right)  \nonumber \\
\psi_2 (z) &=& Ne^{\frac{1}{2}i(k_+-k_-)z}(e^z+e^{-z})^{\frac{1}{2}i(k_++k_-)z}. {(\frac
{e^{-z}}{e^z+e^{-z}})}^{ik_+} \nonumber \\
&.& F\left(\frac{1}{2}ik_+-\frac{1}{2}ik_-+\frac{1}{2}-\gamma,\frac{1}{2}ik_+-\frac{1}{2}ik_-
+\frac{1}{2}+\gamma ;\mid 1+ik_+\mid;\frac{e^{-z}}{e^z+e^{-z}}\right) . \nonumber \\
\eea
For the case of scattering 
\beq
k_+=\sqrt{\epsilon -ve^{2i\mu}}; \ \ k_-=\sqrt{\epsilon -ve^{2i\mu}}
\label{k+-}
\eeq
and the general scattering wavefunction is written as,
\beq
\psi_{scatt} (z)=A\psi_1(z)+B\psi_2(z)
\eeq
where $A$ and $B$ are the arbitrary constants. By considering the asymptotic behavior of $\psi_1$ and 
$\psi_2$ and using the method outlined in Ref. \cite{fl} we construct the left and right handed 
transmission and reflection 
amplitudes as,
\beq
t_l=\frac{1}{G2}; \ \ \ \ \ r_l=\frac{G1}{G2} \ \ \ , 
\label{rtl1}
\eeq
\beq
t_r=\frac{G2.G3-G1.G4}{G2}; \ \ \ \ \ r_r=-\frac{G4}{G2} \ \ \ .   
\label{rtr1}
\eeq
where G1, G2 G3, G4 are calculated in terms of Gamma functions as,
\bea
G1&=& \frac{\Gamma(1-ik_+)\Gamma(ik_-)}{\Gamma(-\frac{1}{2}ik_++\frac{1}{2}ik_-+\frac{1}{2}+
\gamma)\Gamma(-\frac{1}{2}ik_++\frac{1}{2}ik_-+\frac{1}{2}-\gamma)} \ ; \nonumber \\
G2&=& \frac{\Gamma(1-ik_+)\Gamma(-ik_-)}{\Gamma(-\frac{1}{2}ik_+-\frac{1}{2}ik_-+\frac{1}{2}-
\gamma)\Gamma(-\frac{1}{2}ik_+-\frac{1}{2}ik_-+\frac{1}{2}+\gamma)} \ ; \nonumber \\
G3&=& \frac{\Gamma(1+ik_+)\Gamma(ik_-)}{\Gamma(\frac{1}{2}ik_++\frac{1}{2}ik_-+\frac{1}{2}+
\gamma)\Gamma(\frac{1}{2}ik_++\frac{1}{2}ik_-+\frac{1}{2}-\gamma)} \ ; \nonumber \\
G4&=& \frac{\Gamma(1+ik_+)\Gamma(-ik_-)}{\Gamma(\frac{1}{2}ik_+-\frac{1}{2}ik_-+\frac{1}{2}-
\gamma)\Gamma(\frac{1}{2}ik_+-\frac{1}{2}ik_-+\frac{1}{2}+\gamma)} \ . \nonumber \\
\label{gg}
\eea
Now we have the following interesting observations for this non-Hermitian PT-
symmetric system.

\subsubsection{Reciprocity and Handedness}

It is clear from Eq. (\ref{rtl1}), (\ref{rtr1}) and (\ref{gg}) that $r_l\not=r_r$ for both 
real or complex potential. For the Hermitian case $k_+ , k_-$ are real and it is straight 
forward to verify that 
\beq
R_l\equiv \mid r_l\mid ^2=\frac{(G1^*).(G1)}{|G2|^2}=\frac{(G4).(G4^*)}{|G2|^2}=\mid r_r\mid ^2
\equiv R_r \ \ \mbox{and} \ \  T_l=T_r 
\eeq 

Thus we have reciprocity in the case of scattering from a real potential. For the 
PT-symmetric non-Hermitian case 
$k_+ , k_-$ are complex with the properties $k_+^*= k_-$  and $k_-^*=k_+$,
hence $R_l\not=R_r$ even though $T_l=T_r$. This is quite natural as parity is violated
in the case of PT-symmetric potential.
 
We show that at some particular positive energies $R_l=R_r$ even for the non-Hermitian PT-symmetric 
potential. In fact for any parametric range of v and $\mu$ (except $\mu=n\pi$, n=0,1/2,1,3/2,... 
when the potential become either real or indefinite) we find certain discrete energy 
points where $R_l=R_r$. Thus reciprocity is restored even for PT-symmetric non-Hermitian 
potential at some discrete energies. It is clear from the Eq. (\ref{rtl1}) to (\ref{gg})
that reciprocity is restored when $\mid G1\mid ^2=\mid G4\mid ^2$ for the PT-symmetric 
non-Hermitian case. This result is well illustrated with the help of Fig.2 where reciprocity 
is obtained in one or multiple energy values. \\

\includegraphics[scale=0.69]{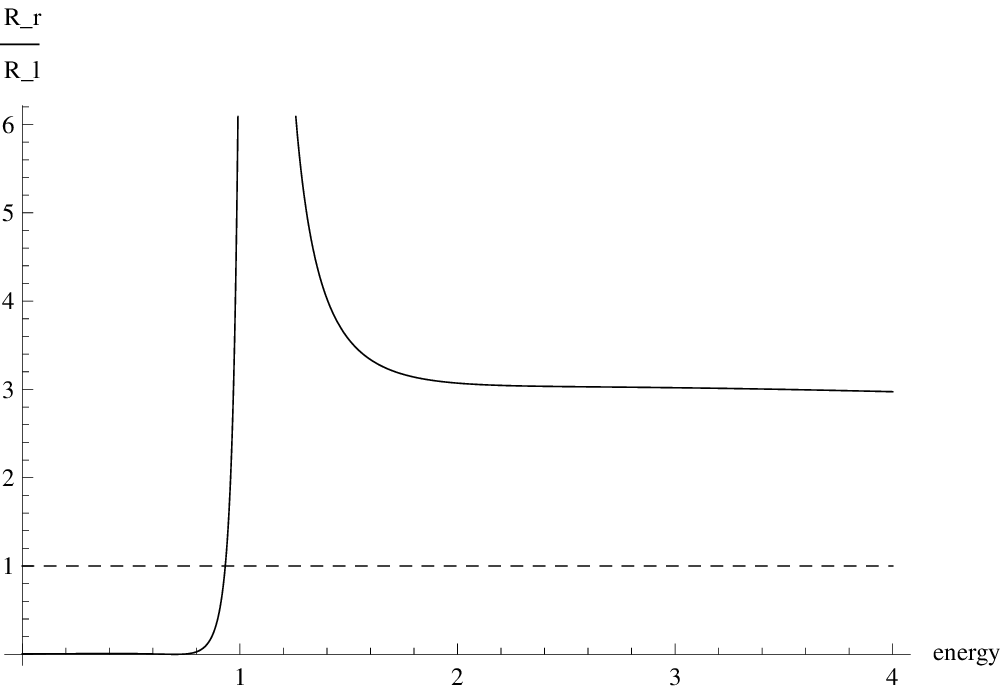} (a) \ \includegraphics[scale=0.69]{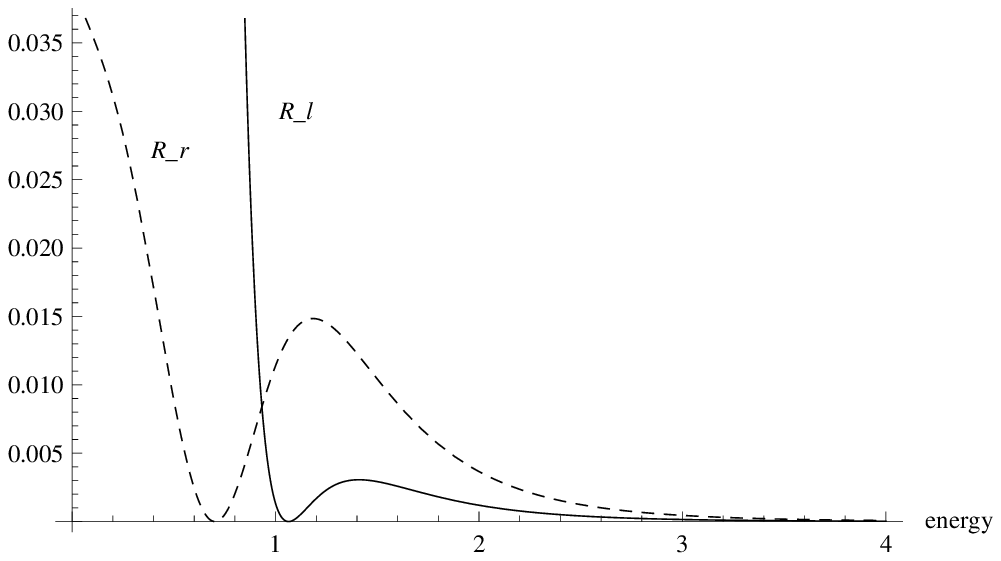} (b) \\

\ \includegraphics[scale=0.7]{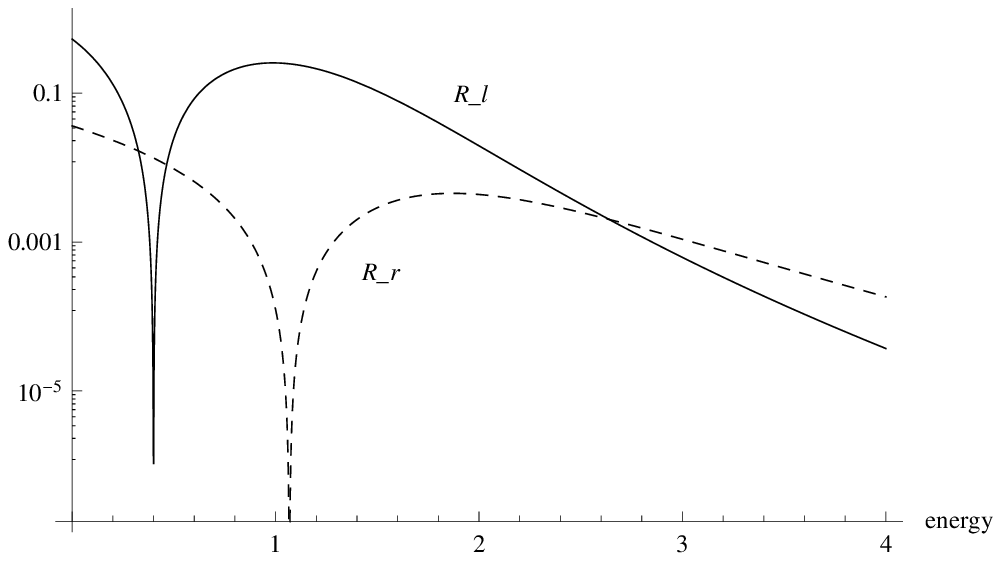} (c) \\

{\bf Fig.2:} {\it Shows different points where reciprocity is restored even for 
PT-symmetric non-Hermitian system. We have one energy point where $R_l=R_r$ for $v=1$ 
and $\mu=\pi/12$, in Fig.2(a). Fig.2(b) shows the left (solid line) and right 
(dashed line) handed reflection coefficients
are equal at the energy point E=0.9943. In Fig.2(c) we have obtained 
three different energy points 0.3255, 0.4642 and 2.639 where reciprocity is restored.}  \\

\subsubsection{Multiple Spectral Singularities (MSS)}

SS are the obstacles for the development of a consistent quantum theory with non-Hermitian 
potential and need to be located with extensive care. We have seen at most one SS in the 
most of 
the PT-symmetric non-Hermitian systems studied in literature so far \cite{ss1, ss2, ss3}. We show 
the existence of
MSS in this PT-symmetric non-Hermitian system. SS arise due to the divergence of the
Gamma functions in the numerator of the reflection and transmission coefficients. The 
divergences occur in $R_l$ ($=\mid r_l\mid ^2$) from Eq. (\ref{rtl1}) due to the Gamma functions in the numerator which
diverges at multiple values of positive energy. This exciting result is demonstrated in
Fig.3. 

We observe another interesting feature of this PT-symmetric non-Hermitian potential. The 
transmission coefficient $T$ oscillates with deep energy minima at the same positive energy 
points where $R_l$ diverges. We note that divergence of the reflection coefficient is dominated over these
deep minima leading to an overall divergence of $R_l$+$T$ in these energy points. The coexistence 
of deep energy minima of $T$ at these MSS points is a unique feature of this PT-symmetric non-Hermitian potential. We further observe that unitarity is restored at certain discrete 
energy values (Fig.3 (b)) and unitarity is violated maximally at the spectral singular points. Unitarity is preserved at sufficient high energy.
\\

\includegraphics[scale=.67]{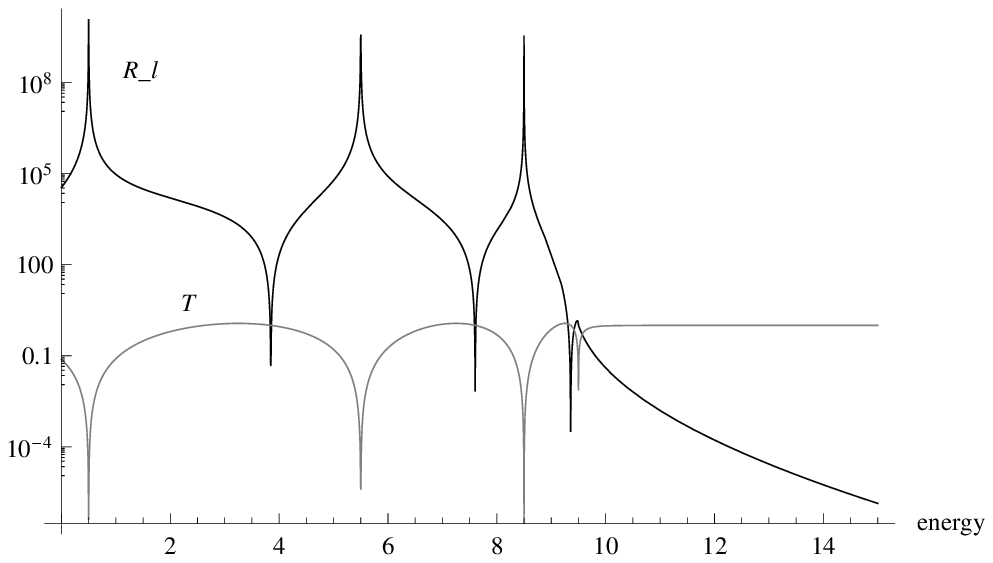} (a) \ \ \includegraphics[scale=.67]{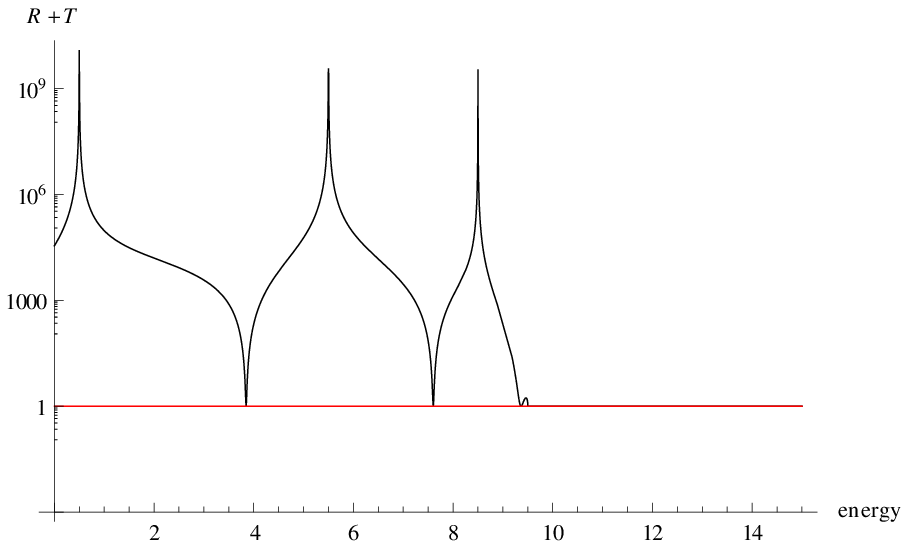} (b) \\

{\bf Fig.3:} {\it The existence of MSS in this non-Hermitian model is shown. In Fig.3(a)
( $v = 9.5, \mu = 6.2832$) the coexistence of divergences of $R_l$ (dark line) and deep energy 
minimas
of $T$ (gray line) are shown. Fig.3(b) shows the multiple spectral singular points where $R_l$+$T$ diverges and 
in the other hand it also
shows the points where unitarity is preserved ($R_l$+$T$=1).}  \\

\subsubsection{Invisibility}

In this section we show that this PT-symmetric non-Hermitian potential has yet another 
interesting property of scattering. The potential become reflectionless for both left and right incidents for certain 
parametric values. However we would like to emphasize that the energy values at which the
potential become reflectionless is different for left and right incidence i.e. E($R_l$=0)
$\not=$ E($R_r$=0). Further we observe that the transmission coefficient becomes unity
at these point. This implies the potential becomes invisible from both sides. However left invisible point is different from right invisible point for this potential. By choosing the
appropriate values of the parameters in the potential it is possible to make this potential
bidirectional invisible at multiple values of energies. All these features are demonstrated in Fig.4. \\

\includegraphics[scale=.65]{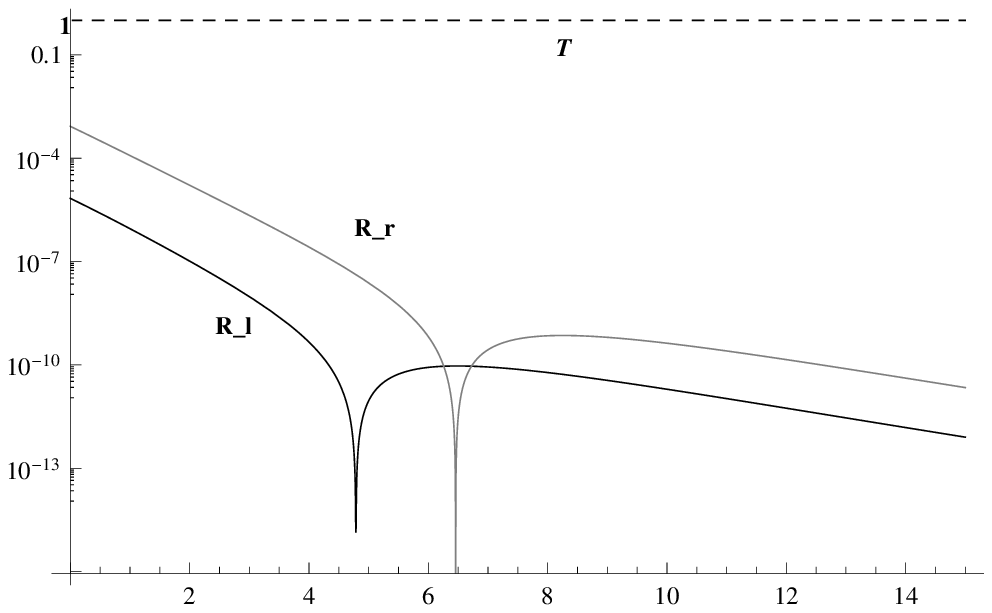} (a) \ \includegraphics[scale=.7]{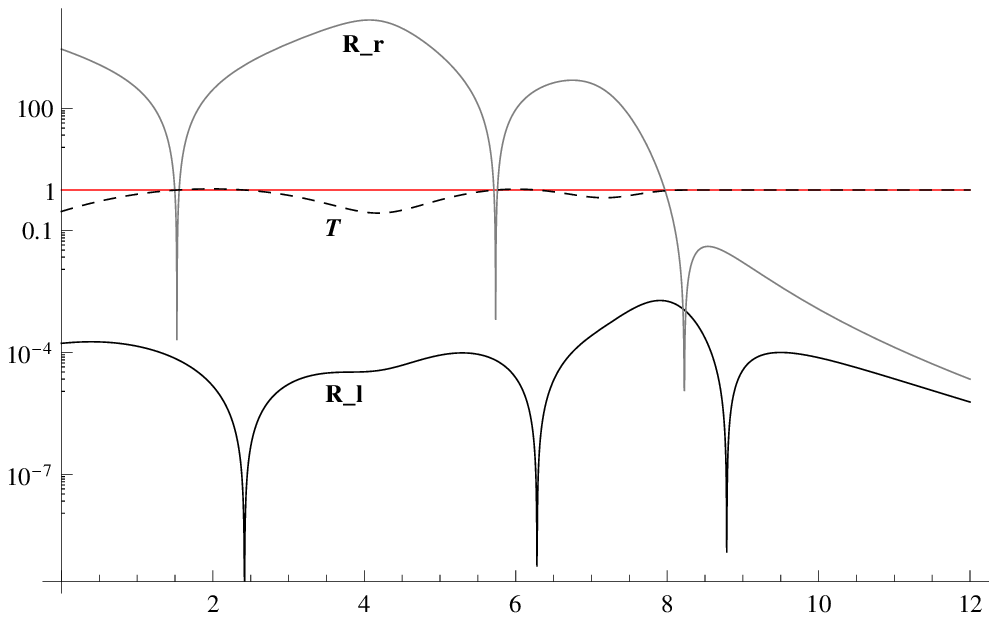} (b) \\

\includegraphics[scale=.7]{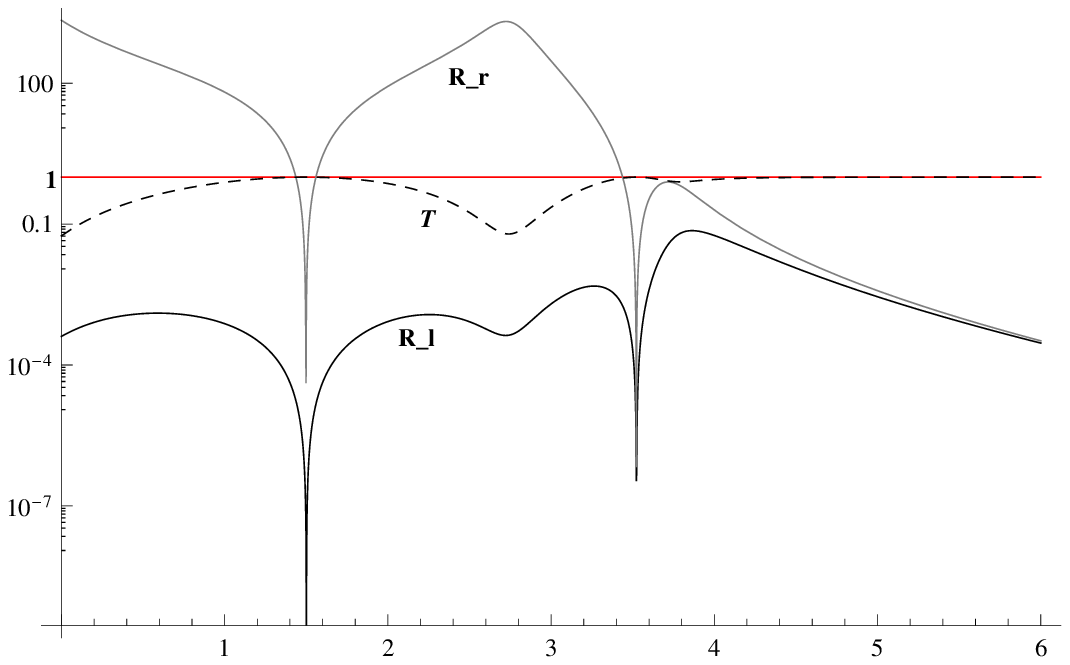} (c)

{\bf Fig.4}: {\it Left (solid line) and right (gray line) handed $R$ and $T$ (dashed line) are plotted against energy to show the 
invisibility ($R$=0, $T$=1) of the potential at discrete energies; In Fig. 4.(a) for $v = 3.54, \mu = 
1.11$ invisible points occurs at different energies for $R_l$ (at E=4.788) and $R_r$ (at 
E=6.455). Fig.4(b) shows multiple invisible points for $v = 8.24, \mu = 6.24$ which are 
unidirectional in nature. Multiple bidirectional invisible points are shown in Fig.4(c) for 
specific values of parameters $v = 3.75; \mu = 3.12$ where $R_l=R_r=0$, and $T$=1 for the same 
energies.}

\section{Non-Hermitian PT-symmetric: case-II ($d\rightarrow id$)}

Time independent Schroedinger equation (TISE) for this case is written as, 
\beq
\frac{\hbar^2}{2 m d^2}\frac{d^2\psi}{dz^2}+V(z)\psi =E\psi
\label{schr}
\eeq
where z is taken as $z=-iX/d-\mu$, with $X=x+i\zeta $.
Note that the differential term in this equation comes with wrong sign due to the presence 
of $d^2$ term.
However this equation can be interpreted as TISE for a upside down potential of the original 
one with energy 
eigenvalues (-E). This upside down potential in the real plane looks as, \\

\ \ \ \ \ \ \ \ \ \ \ \ \ \ \ \ \ \ \ \ \ \ \includegraphics[scale=.9]{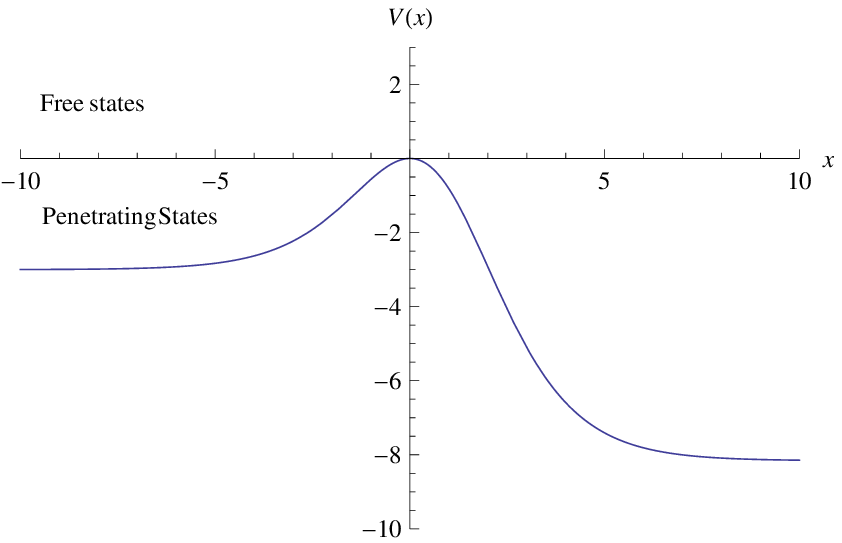}

{\bf Fig.5}: {\it The effective potential (upside down of the original real potential) for
the case $d\rightarrow id$ is plotted for $V_0 = -4.95, \mu = .25$ and d = 2.50.} \\

It behaves like a potential barrier with a maxima at x=0. We have penetrating 
states 
if the particle has negative energy with magnitude less than $ve^{-2\mu}$. On the other hand 
it accepts 
free states when the energy of the particle is more than the barrier height.

The penetrating state solutions are given by,
\bea
\psi_1 (z) &=& Ne^{-az}(e^z+e^{-z})^{-bz} 
F\left(b+\frac{1}{2}-\sqrt{-v\cosh^2 (\mu) +1/4},\right.\nonumber \\
&& \left. b+\frac{1}{2}+\sqrt{-v\cosh^2 (\mu) 
+1/4} ;\mid a+b+1\mid;\frac{e^{-z}}{e^z+e^{-z}}\right)  \nonumber \\
\psi_2 (z) &=& Ne^{bz}(e^z+e^{-z})^{az} 
 F\left(-a+\frac{1}{2}-\sqrt{-v\cosh^2 (\mu) +1/4},\right.\nonumber \\
&&\left. -a+\frac{1}{2}+\sqrt{-v\cosh^2 (\mu) 
+1/4} ;\mid 1-a-b\mid;\frac
{e^{-z}}{e^z+e^{-z}}\right)  \nonumber \\
\label{penn}
\eea
with 
\beq
a^2+b^2=-\epsilon -v\cosh 2\mu; \ \ \ 2ab=-v\sinh2\mu ; 
\eeq
and $k_+=\sqrt{\epsilon +ve^{2\mu}}$, $k_-=\sqrt{\epsilon +ve^{-2\mu}}$
so that 
\beq
a=-1/2(ik_++ik_-); \ \ \ b=1/2(-ik_++ik_-) .
\eeq
The wavefunction for this penetrating state is written in a general form as,
\beq
\psi(z)=A\psi_1(z)+B\psi_2(z)
\label{ppsi}
\eeq
We calculate the left and right handed 
penetrating amplitudes from the asymptotic behavior of Eq. (\ref{ppsi}) as,
\beq
t_l=\frac{1}{P2}; \ \ \ \ \ r_l=\frac{P1}{P2} \ \ \ , 
\label{rtlp}
\eeq
\beq
t_r=\frac{P2.P3-P1.P4}{P2}; \ \ \ \ \ r_r=-\frac{P4}{P2} \ \ \ .   
\label{rtrp}
\eeq
where P1, P2, P3, P4 are expressed in terms of Gamma function written as,
\bea
P1&=& \frac{\Gamma(1-ik_+)\Gamma(-ik_-)}{\Gamma(-\frac{1}{2}ik_+-\frac{1}{2}ik_-+\frac{1}{2}+
\gamma')\Gamma(-\frac{1}{2}ik_+-\frac{1}{2}ik_-+\frac{1}{2}-\gamma')} \nonumber \\
P2&=& \frac{\Gamma(1-ik_+)\Gamma(ik_-)}{\Gamma(-\frac{1}{2}ik_++\frac{1}{2}ik_-+\frac{1}{2}-
\gamma')\Gamma(-\frac{1}{2}ik_++\frac{1}{2}ik_-+\frac{1}{2}+\gamma')} \nonumber \\
P3&=& \frac{\Gamma(1+ik_+)\Gamma(-ik_-)}{\Gamma(\frac{1}{2}ik_+-\frac{1}{2}ik_-+\frac{1}{2}+
\gamma')\Gamma(\frac{1}{2}ik_+-\frac{1}{2}ik_-+\frac{1}{2}-\gamma')} \nonumber \\
P4&=& \frac{\Gamma(1+ik_+)\Gamma(ik_-)}{\Gamma(\frac{1}{2}ik_++\frac{1}{2}ik_-+\frac{1}{2}-
\gamma')\Gamma(\frac{1}{2}ik_++\frac{1}{2}ik_-+\frac{1}{2}+\gamma')} \nonumber \\
\eea
where $\gamma'=\sqrt{-v\cosh^2 (\mu)+1/4} $. Note $P1^*=P4; \ \ P2^*=P3$, as $k_+$ and $k_-$ 
are all real quantities in this case. 
We observe,
$$ r_l\not=r_r \ ; \ \mbox {but} \ R_l\equiv \mid r_l\mid^2=\mid r_r\mid^2\equiv R_r \ \ \mbox {and} \ T_l=T_r \ .$$

This implies we have reciprocity for this non-Hermitian PT-symmetric 
system for all values of energy. However as expected unitarity is violated, i.e $R+T\not=1$ (both for left and right handed cases) for 
this model.
We further observe no SS is present for penetrating states and the potential never becomes 
reflectionless.

\subsection{Free states and Spectral Properties}
We have free state solutions for this case when the energy of the incident particle is 
positive. So the free state solutions of TISE are alike the solutions given
in Eq. (\ref{penn}) with 
\beq
a=1/2(k_+-k_-); \ \ b=1/2(k_++k_-);
\eeq
 where $k_+=\sqrt{\epsilon -ve^{2\mu}}$ and 
 $k_-=\sqrt{\epsilon -ve^{-2\mu}}$ for all positive energy of the incident particle.
The scattering wavefunction is then expressed as,
\bea
\psi(z) &=& \left(e^{-\frac{1}{2}(k_+-k_-)z}(e^z+e^{-z})^{-\frac{1}{2}(k_++k_-)z}\right) 
 \left[ A F\left(\frac{1}{2}k_++\frac{1}{2}k_-+\frac{1}{2}-\gamma, \right.\right.\nonumber \\
&& \left.\left.\frac{1}{2}k_++\frac{1}{2}k_-+\frac{1}{2}+\gamma ;\mid 1+k_+\mid;\frac{e^{-z}}{e^z+e^{-z}}
\right) +B{\left(\frac
{e^{-z}}{e^z+e^{-z}}\right)}^{-k_+} . \right.  \nonumber \\
&&\left. F\left(-\frac{1}{2}k_++\frac{1}{2}ik_-+\frac{1}{2}-\gamma,-\frac{1}{2}ik_++\frac{1}{2}ik_-
+\frac{1}{2}+\gamma ;\mid 1-k_+\mid;\frac{e^{-z}}{e^z+e^{-z}}\right)\right]  \nonumber \\
\eea 
with $z=\frac{x+i\zeta }{i d}-\mu$. Now by considering the asymptotic behavior of the above scattering wavefunction at 
$x, \zeta \rightarrow \pm \infty$ we calculate the left and right handed scattering amplitudes as,
\beq
t_l=\frac{1}{H2}; \ \ \ \ \ r_l=\frac{H1}{H2} \ \ \ , 
\label{rtl}
\eeq
\beq
t_r=\frac{H2.H3-H1.H4}{H2}; \ \ \ \ \ r_r=-\frac{H4}{H2} \ \ \ .   
\label{rtr}
\eeq
where H1, H2, H3, H4 are in terms of Gamma function as,
\bea
H1&=& \frac{\Gamma(1+k_+)\Gamma(-k_-)}{\Gamma(\frac{1}{2}k_+-\frac{1}{2}k_-+\frac{1}{2}+
\gamma)\Gamma(\frac{1}{2}k_+-\frac{1}{2}k_-+\frac{1}{2}-\gamma)} \nonumber \\
H2&=& \frac{\Gamma(1+k_+)\Gamma(k_-)}{\Gamma(\frac{1}{2}k_++\frac{1}{2}k_-+\frac{1}{2}-
\gamma)\Gamma(\frac{1}{2}k_++\frac{1}{2}k_-+\frac{1}{2}+\gamma)} \nonumber \\
H3&=& \frac{\Gamma(1-k_+)\Gamma(-k_-)}{\Gamma(-\frac{1}{2}k_+-\frac{1}{2}k_-+\frac{1}{2}+
\gamma)\Gamma(-\frac{1}{2}k_+-\frac{1}{2}k_-+\frac{1}{2}-\gamma)} \nonumber \\
H4&=& \frac{\Gamma(1-k_+)\Gamma(k_-)}{\Gamma(-\frac{1}{2}k_++\frac{1}{2}k_-+\frac{1}{2}-
\gamma)\Gamma(-\frac{1}{2}k_++\frac{1}{2}k_-+\frac{1}{2}+\gamma)} \nonumber \\
\eea

We observe several interesting features in this case. In one hand $R_l$ diverges at several 
positive energy values leading to the existence of MSS. On the other hand the potential 
becomes reflectionless for the left in a different discrete energy as $R_l=0$. Since at 
the same energy point
the transmission coefficient $T$ becomes unity the potential becomes invisible
from left at that discrete energy.   \\ 

\includegraphics[scale=.67]{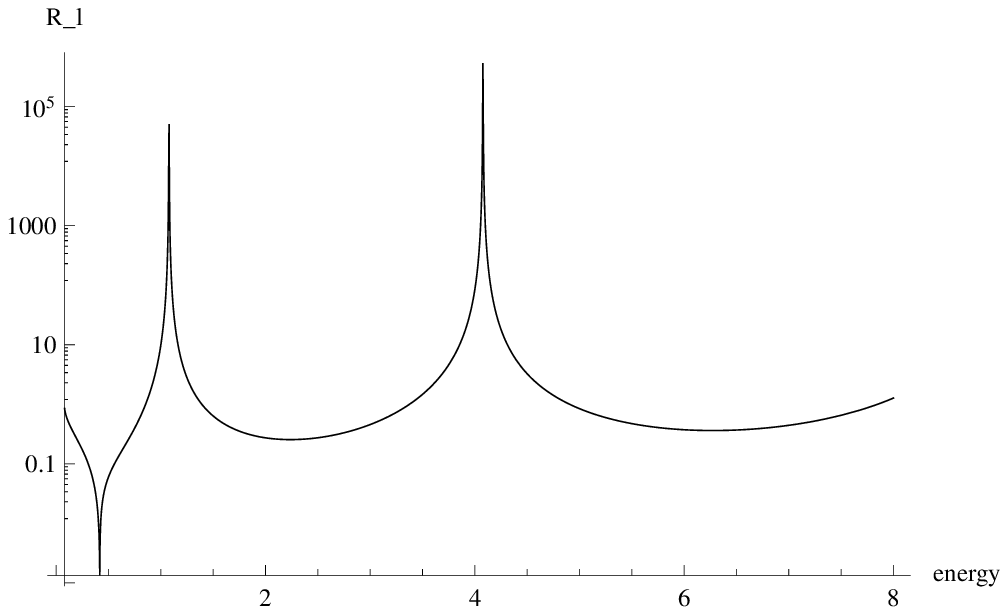} (a) \ \ \ \ \includegraphics[scale=.67]{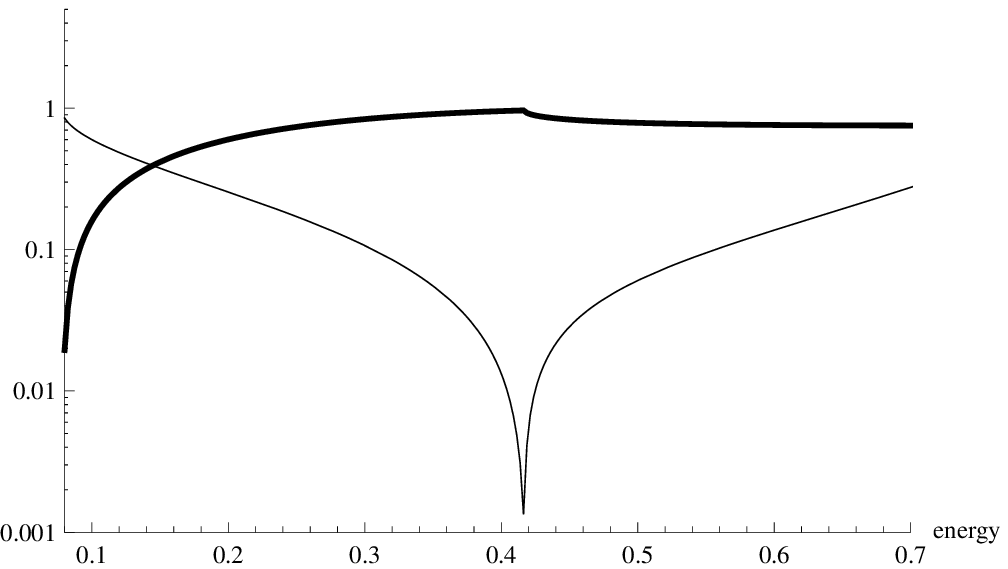} (b) \\

{\bf Fig.6}: {\it In Fig.6(a) spectral singular points are shown where left handed $R$ diverges 
O($10^{6}$) at more than one energy points for specific values of $v$ and $\mu$
($ v = 0.18, \mu =0 .42$). Reflectionless point at E=0.4168 is focused in Fig.6(b) where 
left handed $R$ (dark line) goes to zero and $T$ (bold line) is unity at that energy 
point.} \\

Interestingly we observe no invisible point when the particle 
incidents from right. $R_r$ never becomes zero for this model as none of the argument of 
the Gamma functions in denominator of $R_r$ becomes zero or negative integer. So the potential
in this PT-symmetric configuration shows only unidirectional invisibility for any parametric regime.

\section{Conclusions and Discussions}
We have complexified a one dimensional potential which exhibit bound, reflecting and
free states to capture various interesting properties of non-Hermitian theories. It leads to 
PT-symmetric non-Hermitian system when it is complexified by changing (i) $\mu\rightarrow i\mu$
and (ii) $d\rightarrow id$ . In the first case, when we consider bound states the
system is always in PT-unbroken phase leading to the entire real spectrum. We further  
calculate reflection and transmission amplitudes for both left and right incident particles
by considering the asymptotic behavior of the scattering states. The reflection amplitudes 
are different for left and right incidence both for real as well as PT-symmetric 
non-Hermitian systems (case-I). This implies no-reciprocity for non-Hermitian PT-symmetric 
system. This is natural as reciprocity is a property of a system which remains invariant under
the space inversion (parity). PT-symmetric non-Hermitian potentials essentially change
under parity. However for real potential $R_l=\mid r_l\mid^2=\mid
 r_r\mid^2=R_r$ is always true and we have reciprocity. We have shown that this particular 
non-Hermitian PT-symmetric system is reciprocal at certain discrete energies as shown 
in Fig.2. 

The most interesting and new feature in the scattering of this PT-symmetric non-Hermitian system
is the existence of multiple spectral singularities. This has been shown by considering the 
divergence of $R$ at different positive energies. On the other hand transmission coefficient $T$ 
has deep energy minima 
at these SS points. Coexistence of MSS with deep energy minima 
of transmission coefficient is a unique characteristic of this system. We have further shown
that the potential becomes invisible ($R$=0, $T$=1) at certain energies. It is possible to make 
this potential bidirectional invisible at multiple positive values of energy by adjusting
the parameter of the potential

In the second case ($d\rightarrow id$) the system provides penetrating states if the 
energy of the particle is less than zero and greater than $-ve^{-2\mu}$. We have calculated 
the different transmission and reflection amplitudes both for left and right incidence.
We have analytically shown the reciprocity for this PT-symmetric non-Hermitian 
system. No SS or reflectionless conditions are present for the penetrating states. 
We have also constructed the free states for this case where the energy of the particle 
is positive. We have found the MSS points for this case and the system becomes
reflectionless at certain energies. It will be interesting to restore both reciprocity
and unitarity for non-Hermitian PT-symmetric system at certain energies.

\end{document}